# AutoDeconJ: a GPU accelerated ImageJ plugin for 3D light field deconvolution with optimal iteration numbers predicting

Changqing Su[1,2], Yuhan Gao[3], You Zhou[4], Yaoqi Sun[3], Chenggang Yan[1,3], Haibing Yin[3], Bo Xiong[2,]

[1]School of Mechanical, Electrical&Information Engineering, Shandong University, Weihai 264209, China

[2]National Engineering Laboratory for Video Technology (NELVT), Peking University, Beijing, 100871, China

[3]Lishui Institute of Hangzhou Dianzi University, Hangzhou 323000, China

[4]School of Electronic Science and Engineering, Nanjing University, Nanjing 210023, China

**Abstract**

**Motivation:** Light field microscopy is a compact solution to high-speed 3D fluorescence imaging. Usually, we need to do 3D deconvolution to the captured raw data. Although there are deep neural network methods that can accelerate the reconstruction process, the model is not universally applicable for all system parameters. Here, we develop AutoDeconJ, a GPU accelerated ImageJ plugin for 4.4x faster and accurate deconvolution of light field microscopy data. We further propose an image quality metric for the deconvolution process, aiding in automatically determining the optimal number of iterations with higher reconstruction accuracy and fewer artifacts.

**Results:** Our proposed method outperforms state-of-the-art light field deconvolution methods in reconstruction time and optimal iteration numbers prediction capability. It shows better universality of different light field point spread function (PSF) parameters than the deep learning method. The fast, accurate, and general reconstruction performance for different PSF parameters suggests its potential for mass 3D reconstruction of light field microscopy data.

**Availability and implementation:** The codes, the documentation, and example data are available on an open source at: https://github.com/Onetism/AutoDeconJ.git

## 1 Introduction

The wonder of life's activity lies in its ability to coordinate all its cells and tissues in an elegantly compact system to carry out its functions in an orderly manner. To get a glimpse of this mystery and explore the interrelationships between these parts, various imaging techniques have been proposed gradually, such as two-photon microscopy (Albota et al., 1998), plane illumination methods (Huisken et al., 2004) and confocal microscopy (Schulz et al., 2013), allowing high spatial resolution three-dimensional (3D) imaging (Planchon et al., 2011; Yang et al., 2017; Wu et al., 2021; Xiong et al., 2021). However, much of the interaction between cells and tissues occurs transiently in three dimensions (Prevedel et al., 2014), perhaps in milliseconds or even microseconds. It requires imaging systems with the high spatiotemporal resolution, but the trade-off between space and time can hardly be effectively addressed. Most of the existing techniques prefer to reconstruct a 3D volume by recording a certain number of two-dimensional (2D) images (Keller et al., 2008), which is equivalent to sacrificing temporal resolution for 3D spatial resolution.

Light field microscopy (LFM) has been emerging as a crucial volumetric imaging method due to its ability to capture 3D information in a tomographic manner within a snapshot (Xiong et al., 2021; Cohen et al., 2014). In view of its excellent volumetric imaging speed (Wang et al., 2021), it is exceptionally well suited for high-speed volumetric imaging. As a result, a growing number of biological and medical researchers have paid special attention to applying it in their fields of studies, such as whole-animal 3D imaging of neuronal activity (Prevedel et al., 2014), three-dimensional behavioral phenotyping (Shaw et al., 2018), and high-speed volumetric brain imaging (Zhang et al., 2021). Despite these advantages having led to the rapid development of applications, the presence of post-processing steps for light field images and the low throughput of the reconstruction algorithm at this stage limit its application for long-timescale real-time observation (Wang et al., 2021). 3D Richardson-Lucy (RL) deconvolution algorithm has been widely applied to enhance the resolution of LFM, and Prevedel et al. (2014) have provided software for 3D volume reconstruction in MATLAB based on RL deconvolution (Prevedel et al., 2014). The subsequent related deconvolution methods are also implemented based on RL deconvolution, such as phase-space deconvolution (Lu et al., 2019) and high-resolution light-field microscopy (Li et al. 2019). However, the reconstruction speed of these deconvolution methods is relatively slow, not enough for real-time observation. Although Prevedel et al. (2014) have introduced the GPU acceleration to the deconvolution of light field microscopy data, it is only adopted in part of the convolution operation, thus limiting the overall acceleration performance.

Moreover, it is inconvenient to utilize multiple GPUs simultaneously in MATLAB, which will limit its extension to large-size inputs due to the memory size of the image processor unit (GPU). The deep network XLFMNet (Vizcaino et al., 2021) and VCD-Net (Wang et al., 2021) have been proposed to boost the reconstruction throughput to a fantastic level. However, light field microscopy data with different point spread function (PSF) parameters require training separate specific networks, which makes it trouble for biological researchers since biological observation usually requires different objectives. On the other hand, The trained network is only able to reconstruct the same type of data as the training data. Similarly, the image size in deep learning-based reconstruction is also limited by the memory of the GPU.

Here, to ensure the generality for different system parameters and convenience for users, we design AutoDeconJ, a plugin in ImageJ (Schindelin et al., 2012) for 4.4x faster compared to the Matlab GUI program and accurate deconvolution of light field microscopy data, improving both computational efficiency and convenience of interface interaction. We also add a module to measure the iteration result, predicting the optimal number of iterations. All the main functions of MATLAB versions (Prevedel et al., 2014) are integrated into AutoDeconJ and optimized to take advantage of the parallel processing capacity on the GPU. We first put the time-consuming part of the computation on the GPU, including the part of PSF computing and deconvolution. To maximize the efficiency of parallel computing and solve the problem of insufficient memory on GPU, we also introduce a multi-GPU framework, in which the PSF computation and the reconstruction process of different axial layers in 3D imaging can be evenly distributed to different GPUs, thus doubling the throughput directly. The reconstruction speed is proportional to the number of GPUs in use theoretically. The RL deconvolution algorithm is an iterative process where the number of stop iterations is usually determined by empirical values (Prevedel et al., 2014; Lu et al., 2019). As the input data change, the empirical value may also change, which is highly inconvenient for the reconstruction of large amounts of different light-field data. Hence, we introduce a prediction module that can predict the optimal number of iterations based on the intermediate iterative results. ImageJ is a cross-platform application widely used in biological research (Schindelin et al., 2012), and we thus choose it as our basis in the form of a plugin for all researchers to facilitate their use. We verify the ability of AutoDeconJ in the light field fluorescence data of *C. elegans,* which comes from an open-source dataset (Prevedel et al., 2014). Our developed AutoDeconJ shows its excellent facilitation to light-field reconstruction, including the large data throughput and accurate prediction for iterations. To further demonstrate the performance of AutoDeconJ, we test it on the fluorescence beads data and MCF10A cells data.

## 2 Methods

### 2.1 GPU acceleration

The difference between AutoDeconJ and the previously released MATLAB GUI program (Prevedel et al., 2014) is that AutoDeconJ takes full advantage of the highly parallel computation of the GPU. In the MATLAB GUI program, the calculation of the PSF is all performed on the CPU, which needs a substantial amount of computation. Specifically, the computation time in a single computational cycle is more than millions of microseconds which makes this entire step very time-consuming on the CPU serial data processing model. AutoDeconJ has two significant improvements compared to the MATLAB GUI program. The first improvement is in the acceleration of the calculation of PSF. AutoDeconJ transfers the main time-consuming parts in PSF calculating to the GPU. The speed of PSF calculation can be improved by exploiting the parallel computation (e.g., it achieves more than 20 times improvement in the experiment of *C. elegans*). Another one is our proposed strategy of data processing which is shown in Fig. 1. AutoDeconJ puts most of the operations to GPU during the deconvolution process, and only transfers the final results back to the CPU, avoiding transferring the intermediate results frequently between CPU and GPU, which is time-consuming. In addition, AutoDeconJ also provides support for multi-GPU collaboration to cope with the problem of insufficient memory. After all, not every researcher can afford expensive professional GPUs with large memory. We divide the 3D layers to be reconstructed evenly among different GPUs according to the number of GPUs so that we can handle large-size reconstructions and increase the reconstruction speed exponentially.

### 2.2 Image quality metric

In the field of image processing, the Discrete cosine transform (DCT) has been widely applied to transfer spatial information to the spectral domains, because of its great de-correlation and lossless property (Kristan et al., 2006). The most common mathematical method is to project an image onto an orthonormal basis in which the amplitudes are called the DCT coefficients. Compared to the Discrete Fourier Transform (DFT), the obvious advantage of the DCT is that its coefficients are only represented by real numbers without the complicated complex number operations (Blinn, 1993). Its computing efficiency thus can be significantly improved, making it applicable to some real-time systems. The DCT coefficient transformed from an image function $f(x, y)$ can be obtained via Eq. (1).

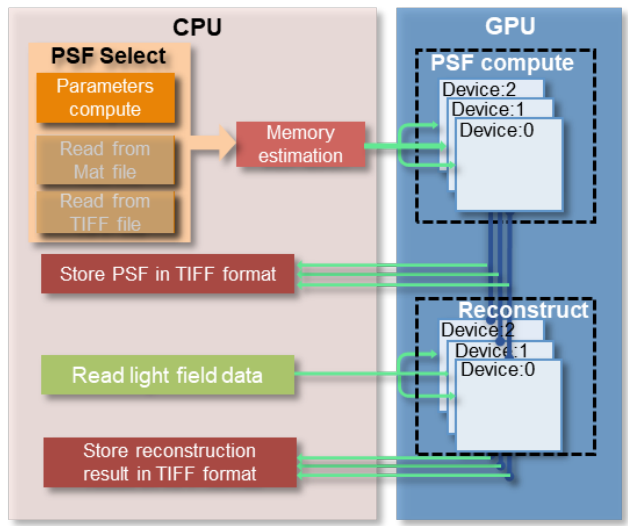

Fig. 1 The acceleration framework of AutoDeconJ. AtuoDeconJ puts the main time-consuming parts, including PSF computation and light-field reconstruction on the GPUs. First, the source of PSF needs to be selected in 3 ways: computing by specific parameters, reading from a MAT file, or reading from a TIFF file. The default is computing by specific parameters. And then, AutoDeconJ will estimate the required memory size to facilitate the following memory allocation for multi-card operation. Whether the PSF is read from a file or calculated based on parameters, the PSF is evenly distributed to each card according to the memory size, which is used directly for the next reconstruction to reduce the time consumption of data transfer between GPU and CPU.

$$F_c(u,v) = \sum_{(x,y)} f(x,y) C_u(x,M) C_v(y,N) \quad (1)$$

where $x \in \{0, \cdots, N-1\}$, $y \in \{0, \cdots, M-1\}$, $M$ is the number of pixels along the height and $N$ is the number of pixels along the width in the image. $C_u(x,M)C_v(y,N)$ is the basis functions defined by the following expressions:

$$C_u(x,M) = c(u,M) * cos(\frac{(2x+1)\pi u}{2M}) \quad (2)$$

$$C_v(y,N) = c(v,N) * cos(\frac{(2y+1)\pi v}{2N}) \quad (3)$$

$$c(\varphi, Z) = \begin{cases} \sqrt{\frac{1}{Z}}; \varphi = 0, \\ \sqrt{\frac{2}{Z}}; otherwise. \end{cases} \quad (4)$$

In general, microscopic observations are made to obtain as much high frequency information as possible. However, due to the limitations of the acquisition system, the resolution information it can obtain is usually in the range of a certain cutoff frequency. We usually pay more attention to information close to the cutoff frequency in the cutoff frequency range rather than near direct current. The relationship between the number of iterations and the image spectrum is illustrated in **Fig. 2a, 2c,** where the source light field image is obtained from the open-source dataset provided in Prevedel et al. (2014). We only take the maximum projection along the z-axis for each iterative result with RL deconvolution.

The deconvolution of the light-filed images can be considered a process of rearrangement of the aliased-signal in the low and high frequency regions, which is reflected in the energy increase in the cutoff frequency range and the extension of effective spectral range in the spectrum of the image. Even the effective spectral range of the reconstructed results will exceed the original cutoff frequency to some extent, due to the additional information introduced by the PSF. Proper deconvolution operations can restore the high frequency detail while maintaining low frequency features. However, excessive deconvolution operations will destroy the original structural information of the image, which is reflected in the DCT transformed image with the periodic spectrum shift, as it has shown in **Fig 2a, 2b, 2c**. During the process of iteration, the maximum amount of information will be presented in and around the cutoff frequency range when the information has been restored to the best. Since the recovery of 3D information is limited, there should be a boundary to the effective spectral region whose size is related to the maximum resolution the optical system can obtain, as shown in **Fig 2c**. A standard metric to measure the region with uniform distribution is the Shannon entropy (Kristan et al., 2006), defined as

$$F_{entropy} = -\sum_{i \in \mathbb{S}} p_i \log_2(p_i) \quad (5)$$

where $p_i$ is a probability function defined on $\mathbb{S}$, and, in our case, $\mathbb{S}$ is selected as a triangular region in the upper left corner in DCT, whose size is defined as

$$G_\mathbb{S} = X_\mathbb{S} * Y_\mathbb{S} / 2 \quad (6)$$

where $X_\mathbb{S}$ is the width of the region, and $Y_\mathbb{S}$ is the height of the region.

For the DCT of an image, assuming that the pixel size is $P_u$, then the spectral range corresponding to the DCT image is $\left[0, \frac{1}{P_u}\right]$. In light field microscopy, $P_u$ can be defined as:

$$P_u = \frac{d_{ML}}{Q * Nnum} \quad (7)$$

where $d_{ML}$ is the microlens pitch size, and $Q$ is the magnification of the objective lens. Assuming that the upper limit of the resolution for the optical system is $d_{psf}$, and its corresponding position $P$ in the DCT image can be defined by:

$$P = \frac{d_{psf}}{P_u} W \quad (8)$$

where $W$ is the pixel number of a dimension (such as $X_\mathbb{S}$ and $Y_\mathbb{S}$ in Eq. (6)) in the DCT image. $X_\mathbb{S}$ and $Y_\mathbb{S}$ can be obtained by:

$$X_\mathbb{S} = P_u * N / d_{psf} \quad (9)$$

$$Y_\mathbb{S} = P_u * M / d_{psf} \quad (10)$$

It is well known that we can estimate the size of the Airy spot from the diffraction theory. For the objective lens, the size of one Airy unit is $\frac{1.22\lambda}{NA}$ (Wang et al., 2010), where $NA$ is the numerical aperture of the objective lens and $\lambda$ is the wavelength of emission light. The above-mentioned $d_{psf}$ is called " Resolution limit for PSF ", which refers to the size of the Airy spot in light field microscopy. Since light field microscopy sacrifices spatial resolution to capture additional angular information, $d_{psf}$ can be defined as (Massaro et al., 2021)

$$d_{psf} = \frac{1.22\lambda * Nnum}{NA} \quad (11)$$

where $Nnum$ is the virtual pixels for each microlens.

The final image metric expression, called DCT entropy, is as follows:

$$DCT_{entropy} = -\frac{2}{\mathbb{S}^2} \sum_{\substack{u<X_\mathbb{S}\\v<Y_\mathbb{S}}} \left|\frac{F_c(u,v)}{L_2(F_c(u,v))}\right| abslog_2\left(\frac{F_c(u,v)}{L_2(F_c(u,v))}\right) \quad (12)$$

where $L_2(\bullet)$ is the 2-norm of a matrix and $F_c(*)$ is DCT which can be referred to Eq.(1) . A more detailed derivation can be found in the supplementary materials. As in **Fig 2d**, we calculate the DCT entropy in the region limited by optical resolution of reconstructed images with different iteration numbers (1~50) and plot the normalized curves accordingly. The results show that the entropy maximum is at the results of the 7th iteration, which is consistent with the empirical value. In practice, we can stop iteration when the DCT entropy value shows a decreasing trend.

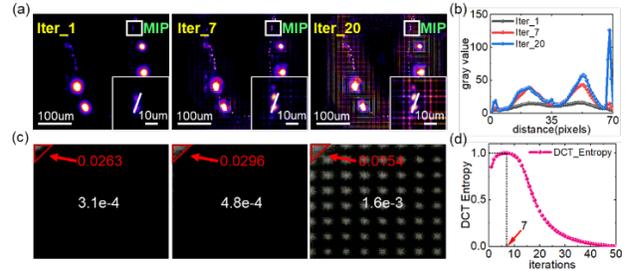

Fig. 2 Performance of the image quality metric. The source light-field data is obtained from the open-source dataset provided in Prevedel et al. (2014). **(a)** The maximum projection along the z-axis for each result with the RL deconvolution at 1, 7, and 20 iterations. The image in the lower-right corner is a magnification of the white area. **(b)** The gray value along solid lines in the lower-right corner of the diagram in (a). **(c)** The DCT transform corresponding to the deconvolution result in (a). The coefficients with higher values are shown in light-gray, and those with lower values are shown in dark-gray. The red numbers are the DCT entropy values of the red triangle in the upper left corner whose size is determined by Eq. (6), and the white numbers indicate the overall DCT entropy value. By visual comparison of the red region as well as the values, it can be observed that with the increase of the iterations, the entropy value in the low-frequency region circled in red will increase first and then become smaller again but the overall entropy keeps rising. **(d)** Normalized

curve of DCT entropy values with the number of iterations for the red region in (c), where the maximum DCT entropy value corresponds to the number of iterations of 7, which is consistent with the empirical value.

## 3   Results

Here we present AutoDeconJ, an ImageJ plugin with a GPU acceleration framework and iterative prediction module for light-field deconvolution. Compared to the specificity and the complicated preparation process of reconstruction by deep learning, the main improvement of AutoDeconJ is to provide a universal tool for light field reconstruction with a decent data throughput, a friendly interactive interface, and the potential to scale to large input sizes. AutoDeconJ is convenient for the user without a background in computer science. In addition, the introduction of the novel iterative criterion in AutoDeconJ can further enhance the user's ability to cope with different kinds of input data.

### 3.1   Availability

Benefiting from ImageJ with powerful cross-platform capability, AutoDeconJ can run on any system with ImageJ or Fiji installed. For details required for ImageJ installation, see the official website *imagej.net* for more information. AutoDeconJ requires the NVIDIA cards support by CUDA8.0 or later. See *https://developer.nvidia.com* for more details about CUDA. If AutoDeconJ needs to run under multiple NVIDIA cards, please ensure that the system is equipped with multiple NVIDIA cards. Our recommendation for these cards is to support the scalable link interface (SLI) or NVLINK, which can further enhance the reconstruction speed. For a detailed user manual, please see supplementary materials. The ImageJ plugin source code is already available on *https://github.com/Onetism/AutoDeconJ.git*. Please clone it to the local folder, then follow the tutorial to compile it, and move the jar package to the */plugin/* folder (where ImageJ is installed) to complete the installation.

### 3.2   Comparison

During development, the main target was to design a universal tool. As such, a final comparison of AutoDeconJ to the Matlab GUI program was performed using the datasets provided in Prevedel et al. (2014), including the fluorescence beads data, and MCF10A cells data. In summary, AutoDeconJ performs as well as the Matlab GUI program in terms of reconstruction quality but requires less time-consuming. It also provides a more friendly interactive interface and a better data throughput. Furthermore, the optimal iteration number predicted by our proposed prediction module is consistent with empirical values, which can be used as a reference for the iterative reconstruction of new light field data. Specific details of the comparison are provided in the supplementary materials. In the end, we also demonstrate the poor data migration capability of the state-of-the-art VCD-Net network on simulated data, which is also presented in the supplementary materials.

## Funding

This work has been supported by Natural Science Foundation of Jiangsu Province, China (No. BK20190292) and National Natural Science Foundation of China under Grant No.62088102.

## References

Albota, M. et al. (1998) Design of organic molecules with large two-photon absorption cross sections. Science, **281**, 1653-1656.

Huisken, J. et al. (2004) Optical sectioning deep inside live embryos by selective plane illumination microscopy. Science, **305**, 1007-9.

Schulz, O. et al. (2013) Resolution doubling in fluorescence microscopy with confocal spinning-disk image scanning microscopy. Proc. Natl Acad. Sci., **110(52)**, 21000–21005.

Keller, P. J. et al. (2008) Reconstruction of zebrafish early embryonic development by scanned light sheet microscopy. Science, **322**, 1065–1069.

Planchon, T. A. et al. (2011) Rapid three-dimensional isotropic imaging of living cells using Bessel beam plane illumination. Nat. Methods, **8**, 417–423.

Yang, W. et al. (2017) In vivo imaging of neural activity. Nat. Methods, **14**, 349–359.

Wu, J. M. et al. (2021) Iterative tomography with digital adaptive optics permits hour-long intravital observation of 3D subcellular dynamics at millisecond scale. Cell, **184**, 3318–3332.e17.

Xiong, B. et al. (2021)Mirror-enhanced scanning light-field microscopy for long-term high-speed 3D imaging with isotropic resolution. Light Sci. Appl., **10**, 227.

Prevedel, R. et al. (2014) Simultaneous whole-animal 3D imaging of neuronal activity using light-field microscopy. Nat. Methods, **11**, 727–730

Wang, Z. Q. et al. (2021) Real-time volumetric reconstruction of biological dynamics with light-field microscopy and deep learning. Nat. Methods, **18**, 551-556.

Cohen, N. et al. (2014) Enhancing the performance of the light field microscope using wavefront coding. Opt. Express, **22**, 24817–24839.

Shaw, M. et al. (2018) Three-dimensional behavioural phenotyping of freely moving C. elegans using quantitative light field microscopy. PloS one, **13(7)**, e0200108.

Zhang, Z. et al. (2021) Imaging volumetric dynamics at high speed in mouse and zebrafish brain with confocal light field microscopy. Nat. Biotechnol, **39**, 74-83.

Lu, Z. et al. (2019) Phase-space deconvolution for light field microscopy. Opt. Express, **27**, 18131–18145.

Li, H. et al. (2019) Fast, volumetric live-cell imaging using high-resolution light-field microscopy. Biomed. Opt. Express, **10(1)**: 29-49.

Vizcaino, J. P. et al. (2021) Real-Time Light Field 3D Microscopy via Sparsity-Driven Learned Deconvolution, 2021 IEEE International Conference on Computational Photography (ICCP), 2021, pp. 1-11.

Schindelin, J. et al.  (2012) Fiji: an open-source platform for biological-image analysis. Nat. Methods, **9**, 676–682.

Kristan, M. et al. (2006) A Bayes-spectral-entropy-based measure of camera focus using a discrete cosine transform. Pattern Recogn Lett, **27**, 1431-1439.

Blinn, J. F. (1993) What's that deal with the DCT? IEEE Comput Graph, **13**, 78-83.

Wang, Y. et al. (2010) Integrated photoacoustic and fluorescence confocal microscopy. IEEE. Trans. Biomed. Eng., **57**, 2576-2578.

Massaro, G. et al. (2021) Light-field microscopy with correlated beams for extended volumetric imaging at the diffraction limit. arXiv preprint arXiv:2110.00807.